# Glassy Relaxation Dynamics in the Two-Dimensional Heavy Fermion Antiferromagnet CeSiI

Kierstin Torres[1†], Joon Young Park[2†], Victoria A. Posey[3], Michael E. Ziebel[3], Claire E. Casaday[4], Kevin J. Anderton[4], Dongtao Cui[4], Benjamin Tang[2], Takashi Taniguchi[5], Kenji Watanabe[6], Abhay N. Pasupathy[7], Xavier Roy[3], and Philip Kim[1,2*]

[1] John A. Paulson School of Engineering and Applied Sciences, Harvard University, Cambridge, MA 02138, USA

[2] Department of Physics, Harvard University, Cambridge, MA 02138, USA

[3] Department of Chemistry, Columbia University, New York, NY 10027, United States

[4] Department of Chemistry and Chemical Biology, Harvard University, Cambridge, MA 02138, USA

[5] Research Center for Materials Nanoarchitectonics, National Institute for Materials Science, 1-1 Namiki, Tsukuba 305-0044, Japan

[6] Research Center for Electronic and Optical Materials, National Institute for Materials Science, 1-1 Namiki, Tsukuba 305-0044, Japan

[7] Department of Physics, Columbia University, New York, NY 10027, United States

[†] These authors contributed equally to this work.

KEYWORDS: CeSiI, heavy fermion, antiferromagnetic order, relaxation dynamics, hysteresis

ABSTRACT: The recent discovery of the van der Waals (vdW) layered heavy fermion antiferromagnetic metal CeSiI offers promising potential for achieving accessible quantum criticality in the two-dimensional (2D) limit. CeSiI exhibits both heavy fermion behavior and antiferromagnetic (AFM) ordering, while the exact magnetic structure and phase diagram have yet to be determined. Here, we investigate magnetic properties of atomically thin CeSiI devices with thicknesses ranging from 2–15 vdW layers. The thickness-dependent magnetotransport measurement reveals an intrinsic 2D nature of heavy fermion behavior and antiferromagnetism. Notably, we also find an isotropic, time-dependent hysteresis in both magnetoresistance and Hall resistance, showing glassy relaxation dynamics. This glassy behavior in magnetic structures may suggest the presence of spin glass phases or multipolar ordering, further establishing CeSiI as an intriguing material system for investigating the interplay between magnetic orders and the Kondo effect.

TEXT: The discovery of van der Waals (vdW) layered materials unlocked numerous opportunities to explore emergent quantum phenomena in a confined phase space[1]. These low-dimensional materials exhibit high sensitivity to thickness, electrostatic gating, and strain, and can be integrated into heterostructures[2]. Furthermore, twisted assemblies of vdW layers can provide additional degrees of freedom to tailor complex phases, including superconductivity[3,4], magnetism[5,6,7], and heavy fermions[8,9]. The recent discoveries of layered heavy fermion materials[10,11] have opened up a plethora of opportunities to explore Kondo physics and quantum criticality in the 2D limit. As an exemplified many-body system, heavy fermions with Kondo interactions are expected to exhibit an emergent phase of matter in the two-dimensional (2D) regime. The competing interplay between the Kondo effect and Ruderman–Kittel–Kasuya–Yoshida (RKKY) magnetic interactions in heavy fermion materials gives rise to rich phase diagrams, often displaying quantum criticality[12,13], unconventional superconductivity[14,15], or non-fermi liquid behavior[16,17]. A systematic study of dimensional cross over between three-dimensional (3D) to 2D regimes may provide further insight into this complex phase, which is the focus our study presented in this paper.

We employ atomically thin CeSiI crystals to carry out thickness-dependent studies. CeSiI is a metallic vdW material, whose unit cell is composed of a silicene layer sandwiched by two layers of triangular lattices of cerium ions and outer iodine layers on each side, terminating the single vdW block[11]. The cerium ions, containing $4f$ orbital electrons with strong electron–electron correlations, supply the localized magnetic moment, a key ingredient for strong correlations. Silicene $sp$ and cerium $d$ orbitals contribute dispersive electrons, which hybridize with the localized moments forming heavy fermions[18,19].

In heavy fermion systems, the competition between magnetic and Kondo interactions is typically driven by non-thermal control parameters such as pressure[14], doping[20], or magnetic field[21,22]. In the previous bulk study[11], CeSiI exhibits signatures of Kondo coupling starting at ~74 K, transitions to the heavy electron phase at lower temperatures, and further transitions to an antiferromagnetic (AFM) phase upon cooling below the Nèel temperature $T_N = 8$ K, all without the application of any non-thermal control parameters. Recent theoretical studies propose that the coexistence of magnetic order and heavy electrons in CeSiI may arise from the decoupling of cerium ions in real space[23] or momentum space[19], allowing some cerium ions to magnetically order while others form heavy fermion states through Kondo interactions.

To date, the exact magnetic structure of CeSiI remains largely unknown. Because of its triangular AFM lattice and the observation of two metamagnetic transitions, CeSiI is presumed to be a frustrated AFM[24]. Powder neutron diffraction studies on CeSiI have not been able to unambiguously determine its magnetic structure, but they suggest cycloid or spin density wave structures as the most probable magnetic ground states[24]. To fully understand the phase diagram of CeSiI and explore quantum criticality and exotic phases, it is crucial to elucidate the details of its magnetic ground state.

In this work, we explore the magnetic properties of CeSiI via electronic transport studies on devices in the atomically thin limit. CeSiI is extremely sensitive to air, necessitating meticulous precautions during device fabrication[11]. We utilize our air- and solvent-free fabrication technique, performing crystal exfoliation, contact fabrication, hBN half-encapsulation, and wire bonding in a glovebox (see Supporting Information section I, Figure S1, and ref. [25] for details). To further protect the device against degradation, we hermetically seal it using a glass slide and low-temperature gallium solder in a pure argon environment. This sealing method significantly improves the stability of devices, enabling us to collect reliable data from atomically thin devices (Figure 1a; see Figure S2 for long-term stability of the sealed device).

Figure 1b shows resistance as a function of temperature, $R(T)$, at zero magnetic field for 2, 4, 8, 15 vdW layers (L) and bulk crystal devices. The thicknesses of the samples are measured using atomic force microscopy after the completion of the transport studies. Across all samples with different thickness, $R(T)$ exhibits two downturns: one at 40 K and another at 8 K corresponding to the onset of the heavy electron Kondo lattice ($T^*$) and antiferromagnetic ordering ($T_N$), respectively, consistent with the previous bulk results[11] without appreciable thickness dependence. These thickness-insensitivity of $T^*$ and $T_N$ suggests the 2D nature of the Kondo lattice and antiferromagnetism in the vdW layered CeSiI, suggesting a minimal interlayer coupling in these orders. This observation further confirms that all devices, down to 2 L, are well protected from environmental degradation. Additionally, we observe a linear relationship between the resistance and $T^2$ below ~25 K across all thicknesses, characteristic of Fermi liquid behavior[11,26] (2 L data shown in the inset of Figure 1b). Note that the quadratic temperature dependence persists below $T_N$, albeit with a much larger slope, potentially supporting the proposal of a heavy Fermi liquid coexisting with a magnetic order in CeSiI[23].

We further explore the thickness effects via magneto-transport measurements. Figure 1c shows the field-symmetrized magnetoresistance (MR; see Supporting Information section I for its definition) as a function of magnetic field $H$ at $T = 2$ K. Consistent with previous results[11], we observe sharp kinks in the MR curve at the two metamagnetic transition fields, $|\mu_0 H_{m1}| \approx 2.5$ T and $|\mu_0 H_{m2}| \approx 4.5$ T, where $\mu_0$ is the permeability of vacuum. The MR rapidly grows at low fields until $H_{m1}$. Following this, the rate of increase slows and the MR plateaus, until reaching $H_{m2}$ where it sharply decreases. After this abrupt drop, the MR continues to increase without saturation. We determine $H_{m1,2}$ for each device by plotting $d(MR)/d(\mu_0 H)$ (Figure S3). We find that the variation of $H_{m1,2}$ is less than 20% as the thickness of the samples vary (inset of Figure 1c), suggesting stable magnetic ground state in the 2D limit, down to the 2 L sample. This thickness-dependent variation is much less than the previous report[11] in which a 4 L CeSiI displayed a considerable reduction in $H_{m1,2}$ up to ~40%. Considering the improved protection from environment using the gallium solder sealing method developed in this study, we attribute the larger variation of $H_{m1,2}$ in the previous study to the increasing disorders in thinner samples due to the degradation of the grease-sealed device, evidenced by its less defined MR features and orders of magnitude larger resistance. We also note that the metamagnetic features in our devices are sharpest for the bulk and 15 L devices yet are still defined down to the 2 L. Additionally, the slope in the MR curve between the two metamagnetic transitions is flatter for thinner flakes and the ratio between metamagnetic features to overall MR background is higher for thinner flakes. These differences can be potentially attributed to the increased surface scattering or disorder induced by substrate interactions, which is more detrimental in thinner flakes as reported in $NbSe_2$[27] and $TaS_2$[28].

The preserved ultra-thin layers of this heavy fermion metal enable us to investigate the Hall effect, which probes the Fermi surface and carrier density, offering insight into the nature of the heavy fermion state[21,29]. The field-anti-symmetrized Hall resistance, $R_{xy}(H)$, is highly nonlinear with $H$, exhibiting kinks at the two metamagnetic transitions below $T_N$ —across the second transition, it undergoes a drastic change in slope from positive to negative ($T = 2$ K data shown in Figure 1d). Figure 1e plots the temperature dependence of the linear portion of the Hall coefficient $R_H = dR_{xy}/d(\mu_0 H)$ in the low-field regime, $|\mu_0 H| \leq$ 0.5 T. As temperature decreases, the sign of $R_H$ changes from negative to positive between $T = 30$ and 40 K across all thicknesses, suggesting that the sign reversal of $R_H$ is accompanied by the Kondo lattice formation. The Kondo hybridization leads to band renormalization, which can result in Fermi surface reconstruction and change in carrier type upon entering the Kondo lattice regime[29]. Meanwhile, $R_H$ exhibits the maximum at $T_N$ and decreases at lower temperatures, ultimately saturating below 2 K. We note that the nonlinear Hall conductance deviates from a simple multichannel conduction model and that skew scattering can significantly influence the Hall effect in Kondo metals[29], necessitating future studies to rigorously elucidate these features in $R_H$.

We further investigate the temperature dependence of magnetotransport properties of CeSiI focusing on the bilayer. In the MR and $R_{xy}$ presented in Figure 2a and 2b, respectively, we observe the aforementioned kinks and plateau associated with metamagnetic transitions below $T_N$. These features become smoother with increasing temperatures and vanish above $T_N$, indicating a correlation to the AFM interaction. While the sign reversal of Hall coefficient occurs at $T^* \approx 40$ K, the negative MR and the nonlinearity in $R_{xy}(H)$ persists up to $T \approx 60$ K $> T^*$ (inset of Figure 2b; see Figure S4 for additional MR data). At higher temperatures, the MR and $R_{xy}$ transition to positive parabolic and linear behavior, respectively.

The central finding of our work appears in the hysteresis in the MR and $R_{xy}$. Figure 2c and 2d show the hysteresis for the corresponding data sets in Figure 2a and 2b, respectively, with hysteresis defined by subtracting backward (BW) magnetic field sweep data from forward (FW) sweep data. Below $T_N$, besides the peaks at the $H_{m1,2}$ (vertical dashed lines) and plateau-like features between them, which we attribute to the two-stage metamagnetic transitions, a broader finite hysteresis is identified in both the MR and $R_{xy}$. Strikingly, this hysteresis remains nonzero even at $H > H_{m2}$, as clearly displayed in the MR hysteresis (inset of Figure 2c, where $\Delta MR = MR^{FW} - MR^{BW}$), while it broadens in field range as the temperature increases to $T_N$. Furthermore, its magnitude is largest at $T = 3$–$4$ K and decreases at lower temperatures. It is in stark contrast to the hysteresis between the two metamagnetic transitions, which monotonically increases and saturates as the temperature decreases. This peculiar temperature- and field-dependence suggests that the broad hysteresis cannot be solely attributed to the AFM ordering, indicating the possible presence of more exotic magnetic phases.

To elucidate the broad hysteresis, we perform hysteresis measurements at different field sweep rates $\mu_0 dH/dt$, where $t$ is time, ranging from 5 to 100 G/s, as shown in Figure 2e and 2f. The broad hysteresis component shows a clear rate dependence: it diminishes at slower sweep rates, largely vanishing at 5 G/s. This again differs from the hysteresis between the $H_{m1}$ and $H_{m2}$, which remain unperturbed by field rate. The observation of time-dependence in hysteresis is evidence for slow relaxation dynamics taking place in CeSiI.

To quantitatively explore the time-dependent nature of the hysteresis, we plot the root-mean-square MR hysteresis $A_{RMS} = \sqrt{\frac{1}{H_R} \int_0^{H_R} (\Delta MR)^2 \, dH}$ for the 2 L and 4 L devices in Figure 3a as a function of $\tau = \left( \frac{1}{H_R} \frac{dH}{dt} \right)^{-1}$, the time taken from zero field to reach $\mu_0 H_R = 7$ T where the hysteresis largely vanishes. We observe a time dependence in which $A_{RMS}$ decreases with increasing $\tau$, following a power law decay (dashed curves in Fig. 3a), further demonstrating slow relaxation dynamics in the CeSiI (see Table S1 for the fitting parameters). We also note that this integrated area includes both the broad time-dependent and time-independent hysteresis contributions. Thus, over slow scan rates and thereby large values of $\tau$, $A_{RMS}$ saturates to a finite value dominated by the time-independent contribution.

We additionally plot $A_{RMS}$ as a function of temperature for field sweep rates of 100 and 10 G/s for the same 2 L and 4 L devices (Figure 3b). Remarkably, we observe a peak in $A_{RMS}$ at $T = 4$ K for both devices at the fast field sweep rate, contrasting to the slow sweep rate in which $A_{RMS}$ is significantly reduced and increases with decreasing temperature below $T_N$ due to the onset of the AFM time-independent hysteresis contribution. The unusual temperature dependent behavior of the broad hysteresis component further substantiates the different origins for this sweep rate dependent hysteresis. This is further demonstrated in the 8 L device taken in the low-temperature regime down to 0.3 K at field sweep rates of 167 G/s and 10 G/s, respectively (Figure S5). At $T = 0.3$ K, we observe the time-dependent contribution vanishes entirely, closely resembling the scan with 10 G/s where only the time-independent contribution remains. The unexpected temperature evolution of the time-dependent hysteresis implies complex magnetic phases present: if the hysteresis strictly originated from typical AFM ordering, we would anticipate an increase in signal with decreasing temperature, as thermal contributions decrease.

The slower relaxation and the increased $A_{RMS}$ for the 4 L device at the fast field sweep rate, relative to the 2 L, imply a dimensional component to the time-dependent hysteresis. Therefore, we investigate the dimensionality of the two hysteresis components through thickness- and field-angle-dependence studies. As both hysteresis components are prominent in the MR, we focus on the MR for the remainder of the hysteresis discussion. At $T = 4$ K, where the time-dependent hysteresis is most prominent, we observe that the time-dependent contribution increases with thickness (Figure 4a and 4b). We note that the MR is normalized with respect to the zero-field longitudinal resistance, and thus the broad hysteresis scaling with the thickness suggests an origin of 3D nature. Figure 4c and 4d present the MR and the corresponding hysteresis of the 4 L device at field angles $\theta = 0°$ and $63°$, where $\theta$ is defined as the angle between $H$ and the $c$-axis of CeSiI. Firstly, the $H_m$ is only sensitive to the out-of-plane component of the magnetic fields $H_\perp = H\cos\theta$ (see Figure 4c), and the hysteresis associated with the metamagnetic transitions at the two field angles also aligns when plotted against $H_\perp$ (Figure S6), supporting nearly collinear 2D AFM structures as reported previously[11]. In contrast, the time-dependent, broad hysteresis component does not vary significantly with $\theta$, as presented in Figure 4d plotted against the total magnetic field $H$. This suggests that the time-dependent hysteresis is isotropic, agreeing with observations from the thickness dependence analysis.

The time-dependent isotropic hysteresis, emerging at $T_N$ together with the time-independent anisotropic AFM phase, points to possible exotic magnetic textures in CeSiI, such as multipolar ordering or a spin glass phase. These scenarios may explain the observed persistence of hysteresis above $H_{m2}$ and the absence of MR saturation in CeSiI even at fields up to 31 T (refs. [11,30]). A recent theoretical study predicts a multipolar magnetic structure in CeSiI with an overall weak magnetic moment due to strong spin–orbit coupling of the Ce $f$-electrons[19]. Such multipolar magnetic orders can yield so-called hidden order states, which are challenging to detect through conventional magnetization measurement techniques[31,32]. Meanwhile, the magnetic frustration in CeSiI (ref. [24]) or disorder in the crystal may yield uncompensated moments which can lead to spin glass behavior[33,34]. This spin glass scenario can explain the unusual temperature dependence of the time-dependent hysteresis component (see Figure 3b): as temperature decreases, the spin fluctuation begins to freeze out, and the dynamic relaxation time becomes too slow compared to the field sweeping rate, at low temperatures $T < 3$–4 K. Although the microscopic origin of the slow relaxation behavior is beyond the scope of this work, our consistent observation across all thicknesses, from 2 L to bulk, suggests it is unlikely due to impurity phases other than CeSiI. Moreover, the more pronounced time-dependent hysteresis in thicker devices (Figure 4b) rules out the possibility that this behavior arises from surface degradation due to air exposure or interactions with the substrate.

Lastly, we explored the possible interplay between the two magnetic textures in CeSiI by performing magnetotransport measurements under field-cooling (FC) conditions. The coexistence of AFM and other magnetic phases, including spin glass, has often yielded an exchange bias[35,36] analogous to what is reported in AFM/ferromagnet heterostructures[37]. In such a picture, upon FC, spins from the glassy or uncompensated phase align with the applied field and if coupled to the AFM phase, pin to the AFM phase. This pinning yields a horizontal shift to the magnetic hysteresis loop, and this offset is known as the exchange bias. We carry out FC and track subsequent hysteresis at various field ranges and temperatures yet have not observed any exchange bias effect in CeSiI (a representative comparison between zero-field-cooling (ZFC) and FC is presented in Figure S7). This observation poses valuable constraints on the possible underlying magnetic phases and their interactions with the AFM order for future studies.

In summary, we demonstrate thickness-dependent magnetic properties of vdW heavy fermion system CeSiI in the atomically thin 2D limit, employing environment-controlled device fabrication processes and hermetic sealing of the devices. By performing magnetotransport study from bulk to bilayer, we confirm the 2D nature of the Kondo lattice formation in a vdW heavy fermion metal CeSiI. We find a two-component hysteresis in the magnetotransport below $T_N$: the first component corresponding to a thickness- and time-independent 2D AFM phase and the other exhibiting a thickness- and time-dependent 3D glassy relaxation behavior. Our work provides a route for further experimental investigation of emergent properties of CeSiI such as the quantum criticality and the interplay between the Kondo effect, antiferromagnetism, and exotic magnetic textures in the 2D limit.

FIGURES:

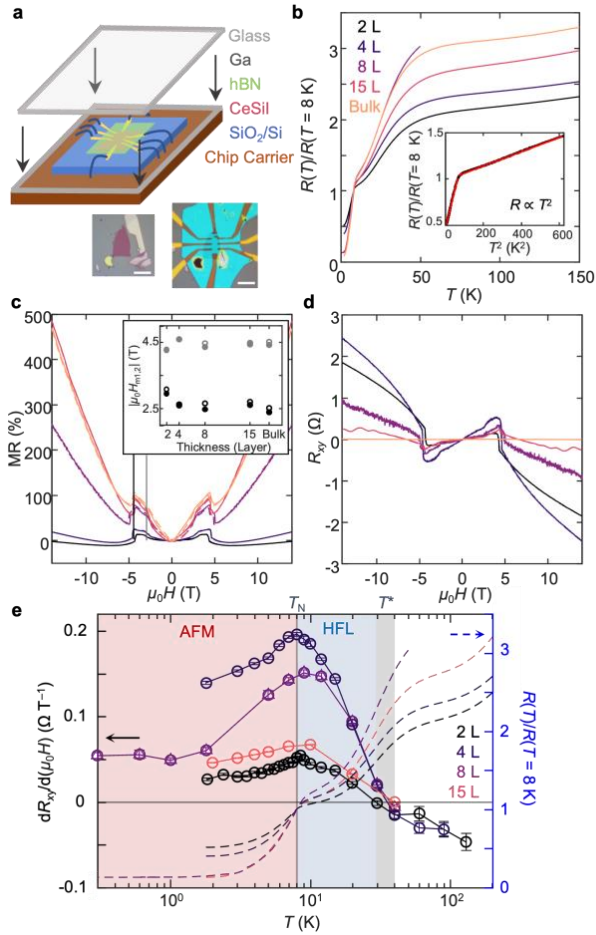

**Figure 1.** (a) Air- and solvent-free fabrication and hermetic sealing of CeSiI devices. Optical images of a 2 L CeSiI exfoliated on SiO₂/Si substrate before (bottom left) and after Au contact deposition and hBN encapsulation (bottom right). Scale bars: 10 μm. (b) Resistance as a function of temperature, $R(T)$, normalized by $R(T = 8\ \text{K})$ for all thicknesses measured. Inset: $R$ versus $T^2$ for 2 L device (black) from 2 to 25 K overlayed with two separate linear fits (red) for the AFM regime (2 to 8 K) and the Kondo lattice regime (8 to 25 K). (c) MR and (d) $R_{xy}$ for all thicknesses at $T = 2$ K. Inset of (c) shows $|\mu_0 H_{m1}|$ (black) and $|\mu_0 H_{m2}|$ (gray) as a function of thickness. Filled (open) markers correspond to data obtained from positive (negative) fields. (e) Low-field Hall coefficient $R_H$ (solid lines with open symbols) overlayed with $R(T)$ (dashed lines) for 2, 4, 8, and 15 L CeSiI devices as a function of temperature. The error bars represent the standard error in the slope of the linear fit to $R_{xy}(|\mu_0 H| \le 0.5$ T$)$. For (c) and (d), solid and dashed lines correspond to forward (FW) and backward (BW) field sweep data, respectively. Inset of (c) and (e) are plotted for FW field sweep. $H \parallel c$ with field sweep rate of 100 G/s for (c–e).

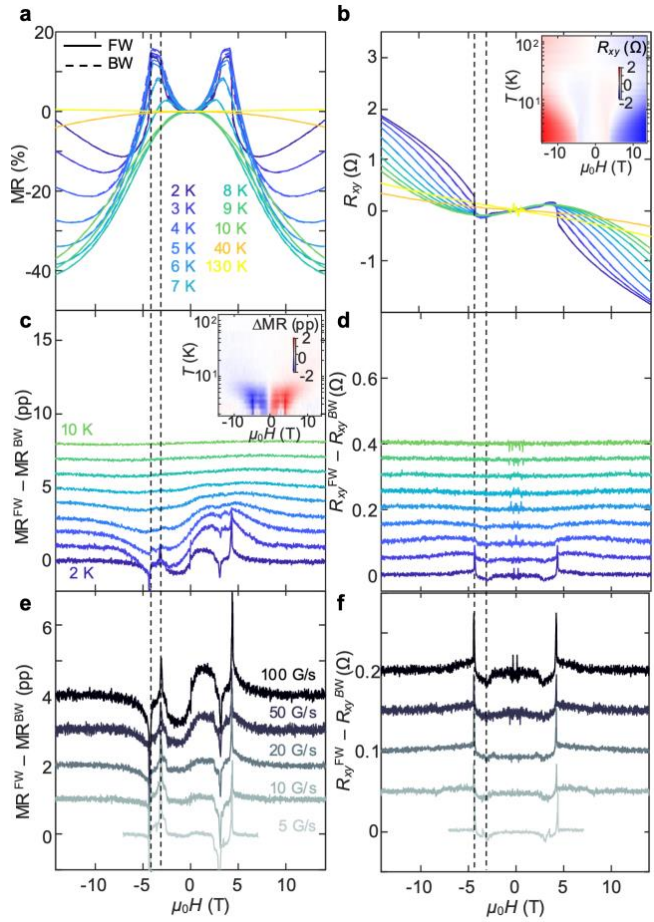

**Figure 2.** Magnetic field dependence of (a) MR and (b) $R_{xy}$ of the 2 L device at different temperatures. FW (BW) field sweeps data are plotted in solid (dashed) lines. Inset in (b): 2D color map of $R_{xy}(H, T)$. (c) MR hysteresis in percentage point (pp), and (d) $R_{xy}$ hysteresis corresponding to (a) and (b), respectively. Inset in (c): 2D color map of hysteresis in MR as a function of $H$ and $T$. Hysteresis in (e) MR and (f) $R_{xy}$ at $T = 2$ K with different field sweep rates. For (a–d), the field sweep rate is 100 G/s. Line plots in (c–f) are shifted vertically for clarity. $H \parallel c$ for all cases. Vertical dashed lines are eye guides for the two metamagnetic transitions. Note that the flux trapped in superconducting magnet and the lagging in lock-in amplifiers with finite time constant can account for artificial hysteresis up to ~200 G.

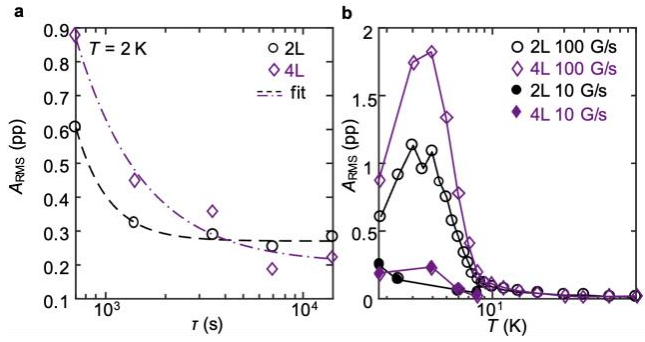

**Figure 3.** (a) Root-mean-square MR hysteresis $A_{RMS}$ for 2 and 4 L devices as a function of $\tau = 7$ T/($\mu_0 dH/dt$) at $T = 2$ K. Dashed lines correspond to power law fits to data with exponents –2.6 and –1.3 for 2 and 4 L devices, respectively (see Table S1 for more detail). (b) $A_{RMS}$ for 2 and 4 L devices as a function of temperature measured with two different field sweep rates of 100 and 10 G/s.

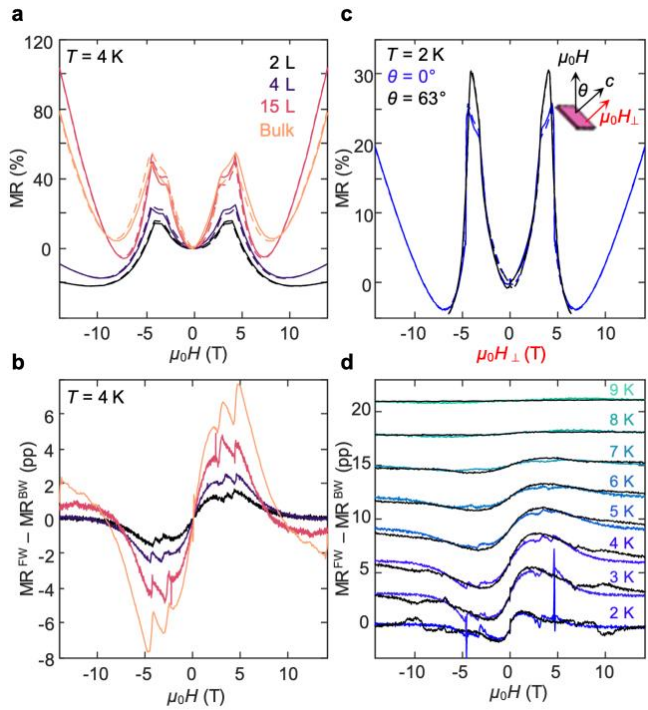

**Figure 4.** (a) MR at $T = 4$ K for different thicknesses and (b) corresponding hysteresis. (c) MR of the 4 L CeSiI for $\theta = 0°$ (blue; $\mu_0 H \parallel c$) and $\theta = 63°$ (black) at $T = 2$ K, plotted as a function of out-of-plane field component $H_\perp$. (d) MR hysteresis corresponding to (c), plotted as a function of total applied field $H$. Each curve, measured at temperatures ranging from 2 to 9 K, is vertically offset by 2.5 percentage points for clarity. For (a) and (c), FW and BW field sweeps data are plotted in solid and dashed lines, respectively. All data are measured with a field sweep rate of 100 G/s.

ASSOCIATED CONTENT

**Supporting Information**:

The following Supporting Information is available free of charge:

Methods, device images, air-stability (sensitivity) of hermetic sealed (unprotected) CeSiI, determination of $H_{m1,2}$, full-temperature MR and $R_{xy}$ data for the 2 L device, power law fitting parameters for $A_{RMS}$, low-temperature MR hysteresis of the 8 L device, additional field-angle-dependent hysteresis plots, and absence of field-cooling dependence in hysteresis.

AUTHOR INFORMATION

**Corresponding Author**

* Email: pkim@physics.harvard.edu

**Author Contributions:** K.T. and J.Y.P. contributed equally to this work.

**Notes:** The authors declare no competing financial interests.

## ACKNOWLEDGMENT

We acknowledge Professor Joseph Heremans for helpful discussions. The major experimental work is supported by NSF (DMR-2105048). Synthesis of CeSiI crystals was supported by the US Department of Energy (DOE), Office of Science, Basic Energy Science, under award DE-SC0023406 (A.N.P., X.R.). K.T., V.A.P., and C.E.C. acknowledge support from the NSF GRFP Fellowship. J.Y.P. acknowledges a partial support from Science Research Center (SRC) for Novel Epitaxial Quantum Architectures (NRF-2021R1A5A1032996). T.T. and K.W. acknowledge support from the JSPS KAKENHI (Grant Numbers 21H05233 and 23H02052) and World Premier International Research Center Initiative (WPI), MEXT, Japan. Nanofabrication was performed at the Center for Nanoscale Systems at Harvard, supported in part by an NSF NNCI award ECCS-2025158.

# Supplementary Information for

# Glassy Relaxation Dynamics in the Two-Dimensional Heavy Fermion Antiferromagnet CeSiI

*Kierstin Torres[1†], Joon Young Park[2†], Victoria Posey[3], Michael E. Ziebel[3], Claire E. Casaday[4], Kevin J. Anderton[4], Dongtao Cui[4], Benjamin Tang[2], Takashi Taniguchi[5], Kenji Watanabe[6], Abhay N. Pasupathy[7], Xavier Roy[3], and Philip Kim[1,2*]*

[1] John A. Paulson School of Engineering and Applied Sciences, Harvard University, Cambridge, MA 02138, USA

[2] Department of Physics, Harvard University, Cambridge, MA 02138, USA

[3] Department of Chemistry, Columbia University, New York, NY 10027, United States

[4] Department of Chemistry and Chemical Biology, Harvard University, Cambridge, MA 02138, USA

[5] Research Center for Materials Nanoarchitectonics, National Institute for Materials Science, 1-1 Namiki, Tsukuba 305-0044, Japan

[6] Research Center for Electronic and Optical Materials, National Institute for Materials Science, 1-1 Namiki, Tsukuba 305-0044, Japan

[7] Department of Physics, Columbia University, New York, NY 10027, United States

[†] These authors contributed equally to this work.

## Table of Contents

Figure S5. Hysteresis in MR measured down to $T$ = 0.3 K

Figure S6. Field-angle dependence of hysteresis in MR

Figure S7. Absence of field-cooling dependence in hysteresis

## III.    References

# I. Methods

**Device fabrication:** All crystal exfoliation, device fabrication, and device packaging are performed in an Ar glovebox with $O_2$ and $H_2O$ concentrations below 0.1 parts per million (ppm). Atomically thin CeSiI flakes are mechanically exfoliated from bulk single crystals onto Si substrates covered with 285 nm $SiO_2$ (see ref.[S1] for details of single crystal growth and bulk physical properties of CeSiI). Prior to exfoliation, the $SiO_2$/Si substrates are cleaned with acetone and isopropyl alcohol and then baked in the glovebox at 300°C for several hours (typically overnight) to remove any adsorbed water on the substrate surface. We screen the exfoliated flakes first by optical microscopy. In order to avoid any exposure to air and solvents, we fabricate contacts via stencil masks patterned on $SiN_x$ membranes (details in ref.[S2]). Once the stencil masks are aligned with the target flakes, we fix the position of stencil mask using Apiezon N vacuum grease. The stencil/sample assemblies are transferred to an evaporator connected to the glovebox. The electrical contacts to CeSiI are formed by electron ($e$)-beam evaporation of 20-nm-thick Au while bonding pads and interconnection lines to the contacts are deposited by another round of $e$-beam evaporation of 3/75 nm Cr/Au (Figure S1). A 10–30 nm thick hBN flake, cleaved from bulk single crystals, is picked up by a polydimethylsiloxane (PDMS) stamp below –90°C and then dropped down over the CeSiI device at room temperature, half-encapsulating the device. The device is cleaved, mounted to a degassed chip carrier, and then wire bonded inside the glovebox. To create the hermetic gallium solder seal, a glass cover slide prewet with gallium is brought into the glovebox and then degassed overnight on a hot plate. The rim of chip carrier is then lined with liquid gallium and capped in the glovebox. Once gallium is frozen (melting point ~30°C), this low-temperature solder seal remains intact even after removal from the glovebox and subsequent pumping (Figure S2). Following measurement, the samples are returned to the glovebox, and we perform atomic force microscopy (AFM) within the glovebox to obtain flake thickness.

Bulk CeSiI devices are fabricated directly on a prebaked chip carrier. The electrical contacts are formed by mechanically pressed indium wires. The bulk devices are hermetically sealed with gallium solder in the same manner.

**Electronic transport measurements:** Magnetotransport measurements are performed with a standard lock-in technique. An AC excitation current $I = 1$ μA (0.1–0.5 mA) at 17.777 Hz is biased to the atomically thin (bulk) CeSiI devices using a lock-in amplifier (Stanford Research Systems SR830) and a current-limiting resistor. The voltage signals are pre-conditioned with low-noise voltage preamplifiers (Stanford Research Systems SR560 and DL Instruments DL1201). The temperature, $T$, and magnetic field, $H$, are controlled using a $^4$He variable temperature insert (Quantum Design Physical Property Measurement System) and a $^3$He insert (Oxford Instruments Heliox VL) equipped with superconducting solenoid magnets. The wires in the latter system are filtered by Mini-Circuits radio frequency low-pass filters and one-pole RLC low-pass filters at cryogenic temperatures. The magnetic field is oriented parallel to the *c*-axis of CeSiI crystals and swept at a rate of 100 G/s unless otherwise stated. The field-angle-dependent study is performed by rotating the sample *ex-situ*.

A typical Hall bar (van der Pauw) geometry is employed for the atomically thin (bulk) devices (measurement configuration indicated in Figure S1d). Due to the irregularities in crystal shapes, a finite mixing between the longitudinal voltage, $V_{xx}$, and Hall voltage, $V_{xy}$, contributions is unavoidable. Therefore, we symmetrize and anti-symmetrize longitudinal and Hall voltage values $V_{xx}(H)$ and $V_{xy}(H)$ to obtain $R_{xx}(H)$ and $R_{xy}(H)$, respectively, by the following expressions:

$$R_{xx}^{\mathrm{FW(BW)}}(H) = \frac{V_{xx}^{\mathrm{FW(BW)}}(H) + V_{xx}^{\mathrm{BW(FW)}}(-H)}{2I},$$

$$R_{xy}^{\mathrm{FW(BW)}}(H) = \frac{V_{xy}^{\mathrm{FW(BW)}}(H) + V_{xy}^{\mathrm{BW(FW)}}(-H)}{2I},$$

where the superscript FW (BW) refers to the field sweep direction from negative (positive) to positive (negative) fields. The magnetoresistance (MR) is calculated by normalizing the field-modulated component by the zero-field value:

$$\text{MR}^{\text{FW(BW)}}(H) = \frac{R_{xx}^{\text{FW(BW)}}(H) - R_{xx}^{\text{FW(BW)}}(H=0)}{R_{xx}^{\text{FW(BW)}}(H=0)} \times 100.$$

We note that the lack of ability to well-define the channel geometry due to the environmental sensitivity of CeSiI results in invasive contact configuration with ill-defined geometric factors, prohibiting quantitative analysis of the Hall effect between the devices in a consistent manner. Nevertheless, the qualitative evolution of the Hall effect with temperature and magnetic field is valid.

## II. Supplementary Figures

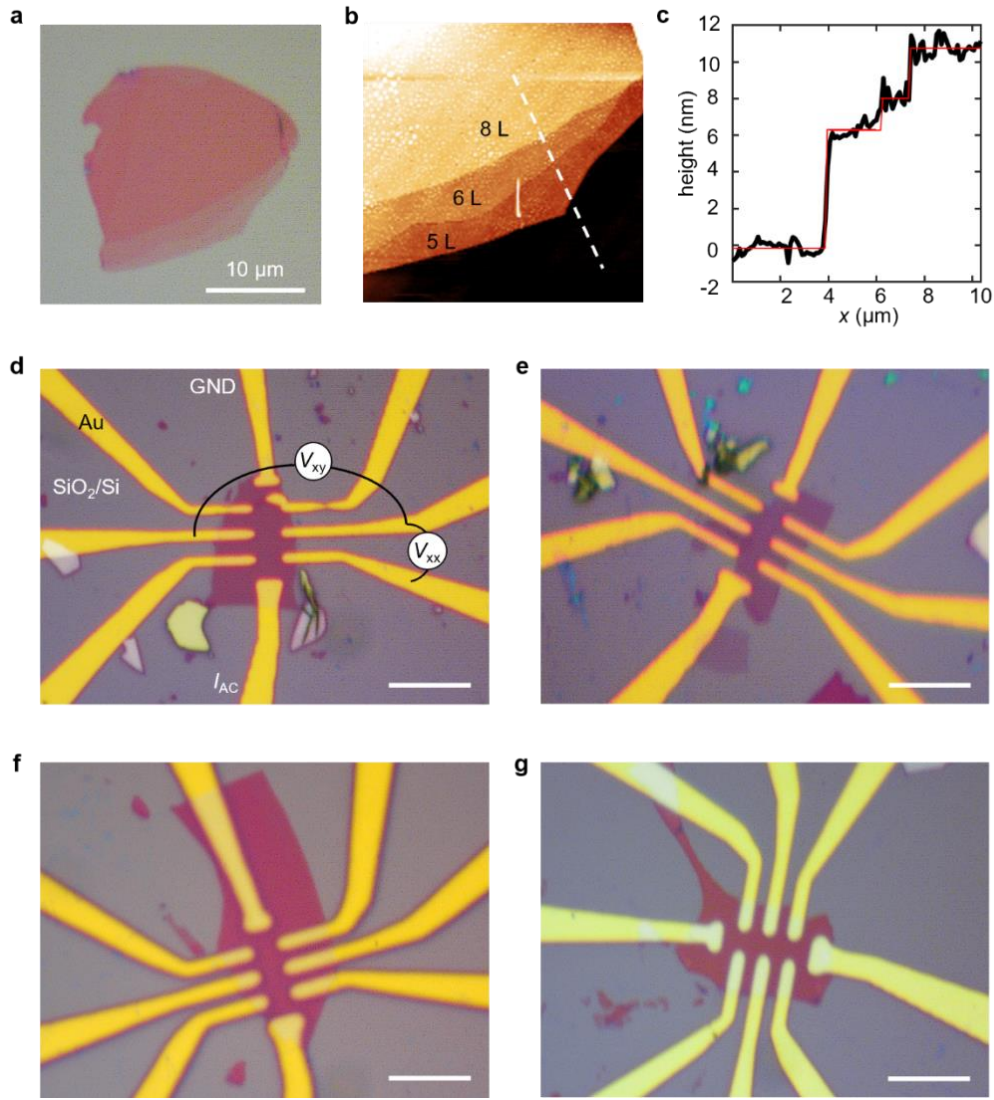

**Figure S1. Optical microscope images of CeSiI devices.** (a) Optical microscope image and (b) AFM topography image of an CeSiI flake exfoliated on a SiO₂/Si substrate, showing the correlation between optical contrast and thickness. Bubbles appear during the AFM scan when the circulation is switched off to minimize vibration. (c) AFM height profile along the white dashed line in (b). Red line is a step function fit to data (black). Optical microscope images of atomically thin CeSiI devices used in magnetotransport studies (captured prior to hBN capping): (d) 2 L, (e) 4 L, (f) 8 L, and (g) 15 L device. Scale bars: 10 μm. Typical measurement configuration is overlayed in (d).

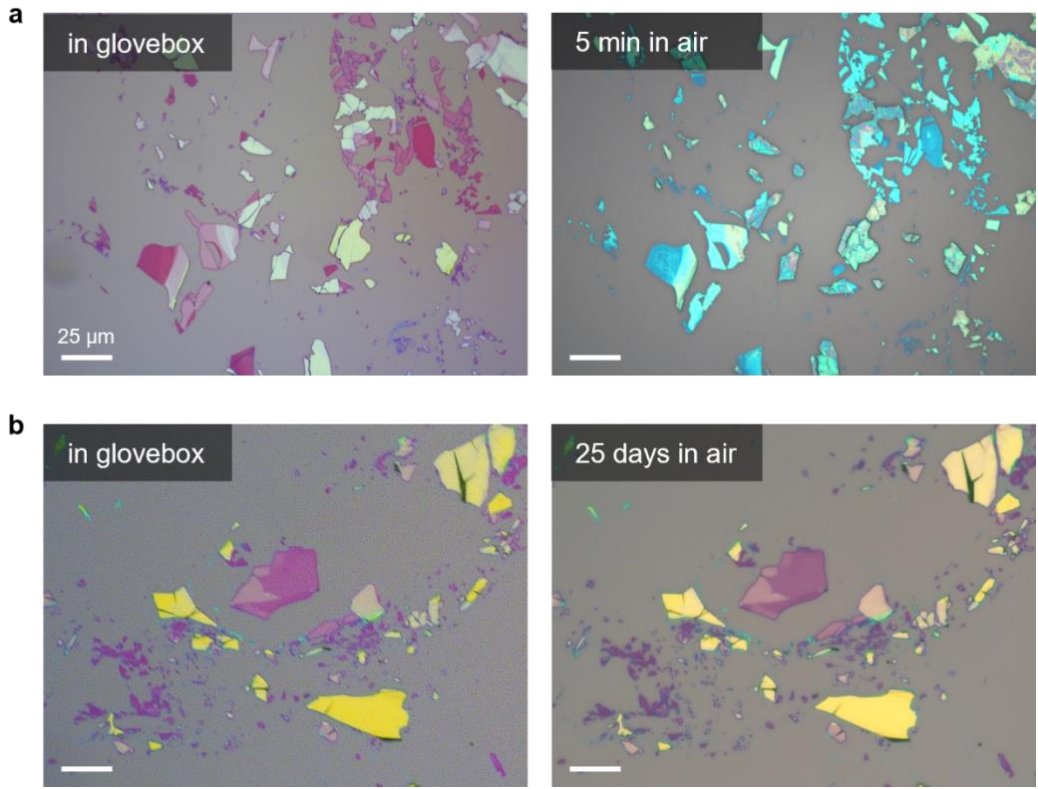

**Figure S2. Air sensitivity of CeSiI and stability of hermetical seal.** (a) Optical microscope images of exfoliated CeSiI flakes in the glovebox (left) and after 5 minutes of air exposure without any protection (right). Changes in flake color and bubbles formation on the surface occur immediately upon air exposure. (b) Flake stability with gallium solder seal. Optical microscope images of exfoliated CeSiI flakes in the glovebox (left) and after 25 days in air while sealed in the chip carrier with gallium solder (right; image taken through a cover glass). Scale bars: 25 μm. No significant change to flake color or signs of degredation are observed. Between capturing the left and right images, the package has been subjected to three cycles of vacuum pumping and venting to the atmosphere to further test the stability of the hermetic seal.

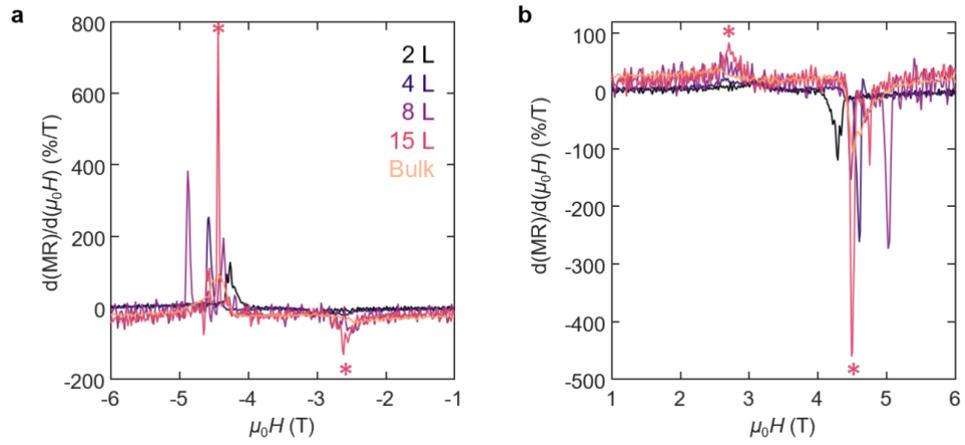

**Figure S3. Metamagnetic transition fields.** d(MR)/d($\mu_0H$) at (a) negative and (b) positive fields. We define the metamagnetic transition field as the field where the peak in d(MR)/d($\mu_0H$) occurs (shown by the * symbol for the 15 L device). FW field sweep data is shown.

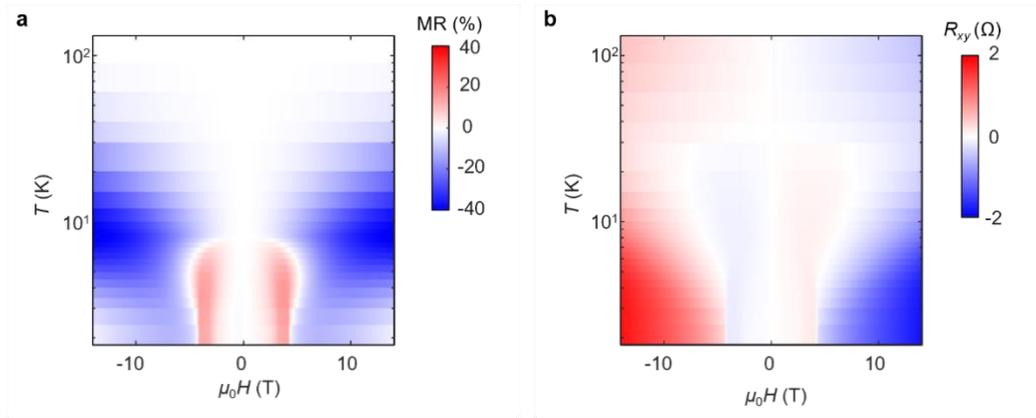

**Figure S4. Additional magnetotransport data.** 2D color plot of (a) MR($H$, $T$) and (b) $R_{xy}$($H$, $T$) of the 2 L device across the full temperature range (1.8 to 130 K). FW field sweep data is shown.

| Device | $A_0$ | $b$ | $a$ |
|--------|-------|-----|-----|
| 2 L | 0.27 | $1.06 \times 10^7$ | $-2.64$ |
| 4 L | 0.20 | $2.93 \times 10^3$ | $-1.28$ |

**Table S1. Power law fit results for root-mean-square MR hysteresis.** Values of $A_0$, $b$ and $a$ for the power law fits $A_{\text{RMS}} = A_0 + b\tau^a$ displayed in Figure 3a, where $A_{\text{RMS}} = \sqrt{\frac{1}{H_{\text{R}}} \int_0^{H_{\text{R}}} (\Delta\text{MR})^2 \, \text{d}H}$ is the root-mean-square MR systeresis, and $\tau = 7$ T/($\mu_0\text{d}H/\text{d}t$).

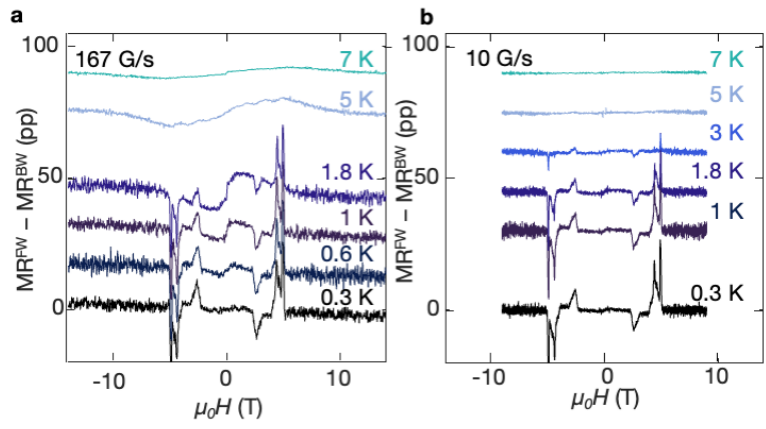

**Figurre S5. Hysteresis in MR measured down to $T$ = 0.3 K.** MR hysteresis versus $\mu_0H$ for the 8 L device, taken with a field sweep rate of (a) 167 G/s and (b) 10 G/s at different temperatures. The broad hysteresis component vanishes at 0.3 K, while the hysteresis between two metamagnetic transitions becomes more pronounced as temperature decreases.

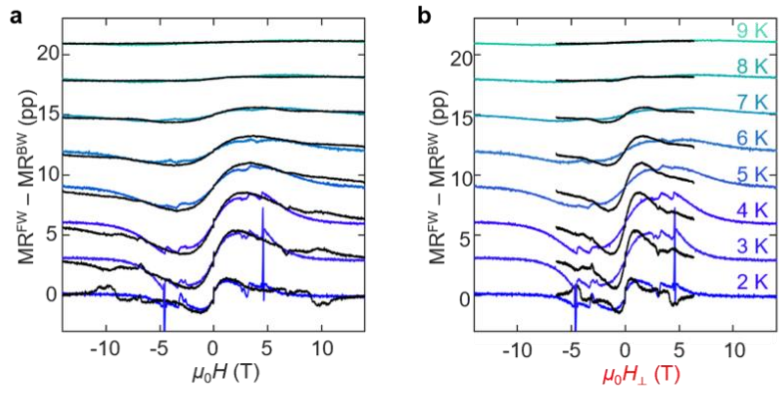

**Figure S6. Field-angle dependence of hysteresis in MR.** MR hysteresis of the 4 L device at field angle $\theta = 0°$ (black) and $\theta = 63°$ (blue) with respect to the $c$-axis of CeSiI, ploted versus (a) total applied field $H$ and (b) out-of-plane field component $H_\perp$. Each curve, measured at a fixed temperature ranging from 2 to 9 K, is vertically offset by 2.5% for clarity. The broad hysteresis component is insensitive the the field angle, while the hysteresis associated with the two-stage metamagnetic transitions at both field angles aligns when plotted against $H_\perp$.

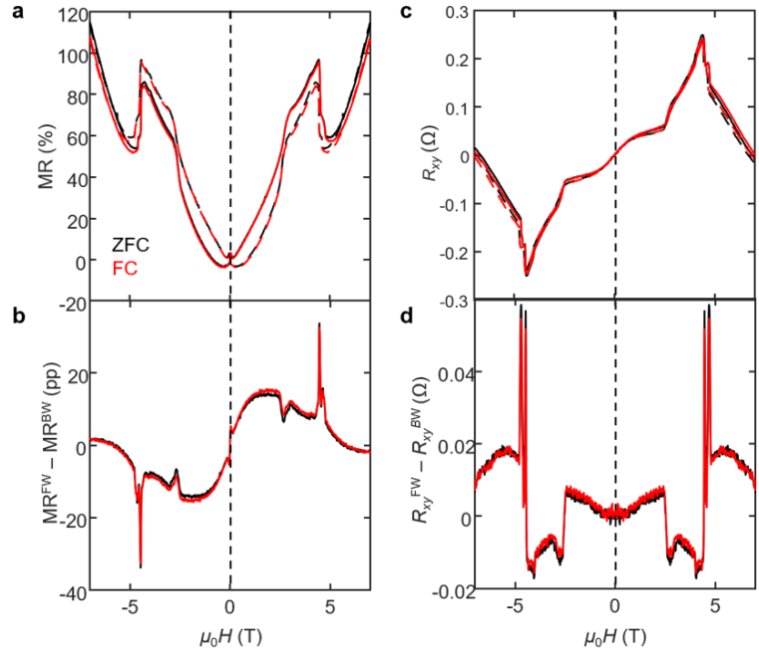

**Figure S7. Absence of field-cooling dependence in hysteresis.** (a) MR, (b) hysteresis in MR, (c) $R_{xy}$, and (d) hysteresis in $R_{xy}$ for the 15 L device at $T$ = 2 K. The device is ZFC (black) and FC from 60 K to 2 K with $\mu_0 H$ = 1 T (red). The field is subsequently ramped to 7 T, and then BW (dashed) and FW (solid) field sweeps are performed. The hysteresis does not shift in field upon FC, suggeting absence of the exchange bias effect.